\documentclass[usenatbib,epj,twocolumn]{webofc}
\usepackage[varg]{txfonts}   
\usepackage[rightcaption]{sidecap}
\providecommand{\xray  }{X-ray\xspace}%
\providecommand{\gray  }{$\gamma$-ray\xspace}%
\providecommand{\grays }{$\gamma$-rays\xspace}%
\providecommand{\fermi }{\textit{Fermi}\xspace}%
\providecommand{\swift}{\textit{Swift}\xspace}%
\providecommand{\planck}{\textit{Planck}\xspace}%
\providecommand{\wise}{\textit{Wise}\xspace}%
\providecommand{\herschel}{\textit{Herschel}\xspace}%

\usepackage{colordvi}
\woctitle{The Innermost Regions of Relativistic Jets and Their Magnetic Fields}
\begin{document}
\title{Magnetic Field Amplification and Blazar Flares}
%
%

\author{Xuhui Chen\inst{1,2}\fnsep\thanks{\email{xuhui.chen@alumni.rice.edu
    }} \and
        Ritaban Chatterjee\inst{3} \and
        Giovanni Fossati\inst{4} \and
        Martin Pohl\inst{1,2}
}

\institute{Institute of Physics and Astronomy, University of Potsdam, 14476 Potsdam-Golm, Germany 
\and
           DESY, Platanenallee 6, 15738 Zeuthen, Germany 
\and
	   Department of Physics, Presidency University, 86/1, College street, Kolkata-700073, India
\and
           Department of Physics and Astronomy, Rice University, Houston, TX 77005, USA
          }

\abstract{%
  Recent multiwavelength observations of PKS 0208-512 by SMARTS, \fermi, and \swift revealed that \gray 
and optical light curves of this flat spectrum radio quasars 
are highly correlated, but with an exception of one large 
optical flare having no corresponding gamma-ray activity or even detection. 
On the other hand, recent advances in SNRs observations and plasma simulations both reveal that 
magnetic field downstream of astrophysical shocks can be largely amplified beyond simple shock compression. 
These amplifications, along with their associated particle acceleration, might contribute to 
blazar flares, including the peculiar flare of PKS 0208-512. 
Using our time dependent multizone blazar emission code, we evaluate several scenarios that 
may represent such phenomena. This code combines Monte Carlo method that tracks the radiative processes 
including inverse Compton scattering, and Fokker-Planck equation that follows the cooling and 
acceleration of particles. It is a comprehensive time dependent code that fully takes into account 
the light travel time effects. In this study, both the changes of the magnetic field and 
acceleration efficiency are explored as the cause of blazar flares. Under these assumption, 
synchrotron self-Compton and external Compton scenarios produce distinct features that 
favor the external Compton scenario. 
The optical flares with/without gamma-ray counterparts can be explained by 
different allocations of energy between the magnetization and particle acceleration, 
which in turn can be affected by the relative orientation between the magnetic field and the shock flow. 
We compare the details of the observations and simulation, and highlight what implications 
this study has on our understanding of relativistic jets. 
}
\maketitle
\section{Introduction}
\label{intro}
As an extreme class of Active Galactic Nuclei (AGNs), blazars are known to emit electromagnetic waves 
in almost all frequencies that are currently being observed, 
extending from radio to \gray. They are also famous
for being highly variable in an unpredictable manner. An immediate interesting question people begin to ask
is whether blazar variations in different frequencies are correlated. If yes, how. For example,
whether there are any time lags, what the amplitude relations are. 
The answers to these questions can identify important physical origins of
those emission, including its location and mechanism. For one type of blazars, namely, flat spectrum radio 
quasars (FSRQs), the correlation between optical emission and GeV \grays is particularly interesting. 
These two energy bands represent the energies either at or beyond the peaks of the two components of 
their spectral energy distributions (SEDs). The photons in these energies are probably emitted 
by the most energetic particles, and hence exhibit most violent variations. 
Identification of these correlation became possible following the
launch of \fermi, as well as the implementation of its supporting optical monitoring programs, 
such as the Yale/SMARTS program.
In most cases the correlation between these two bands are established 
\cite{bonning_2012:smarts_blazar,chatterjee_2012:smarts_similarity}.
However, \cite{chatterjee_2013:0208_opticalonly} identified at least one case of such correlation breaking
down (see also Fig.\,2 of \cite{hess_2013:1510}). 
In contrast to the "orphan" \gray flares occasionally found in BL lac objects 
\cite{krawczynski_etal:2004:1es1959},
these FSRQs show strong optical flares without \gray counterparts. 
The authors of \cite{chatterjee_2013:0208_opticalonly} identified
three major optical flares from PKS 0208-512, with highly correlated \gray activity in flares 1 and 3.
But in flare 2, \gray remains at a low level. 
The question arises as why the same source exhibits correlated optical/\gray flares sometimes, but
orphan optical flares at other times.

\begin{SCfigure*}[0.3]
\centering
\caption{The \fermi (upper panels) and SMARTS (lower panels) optical-near infrared data in 10 day bins. The left and right figures
are flares 1 and 3 identified by \cite{chatterjee_2013:0208_opticalonly}, starting from
Modified Julian Date (MJD) 54680 and 55680.
\newline
\newline}\label{fig:obs_lc}

\hfill
\includegraphics[width=0.68\linewidth]{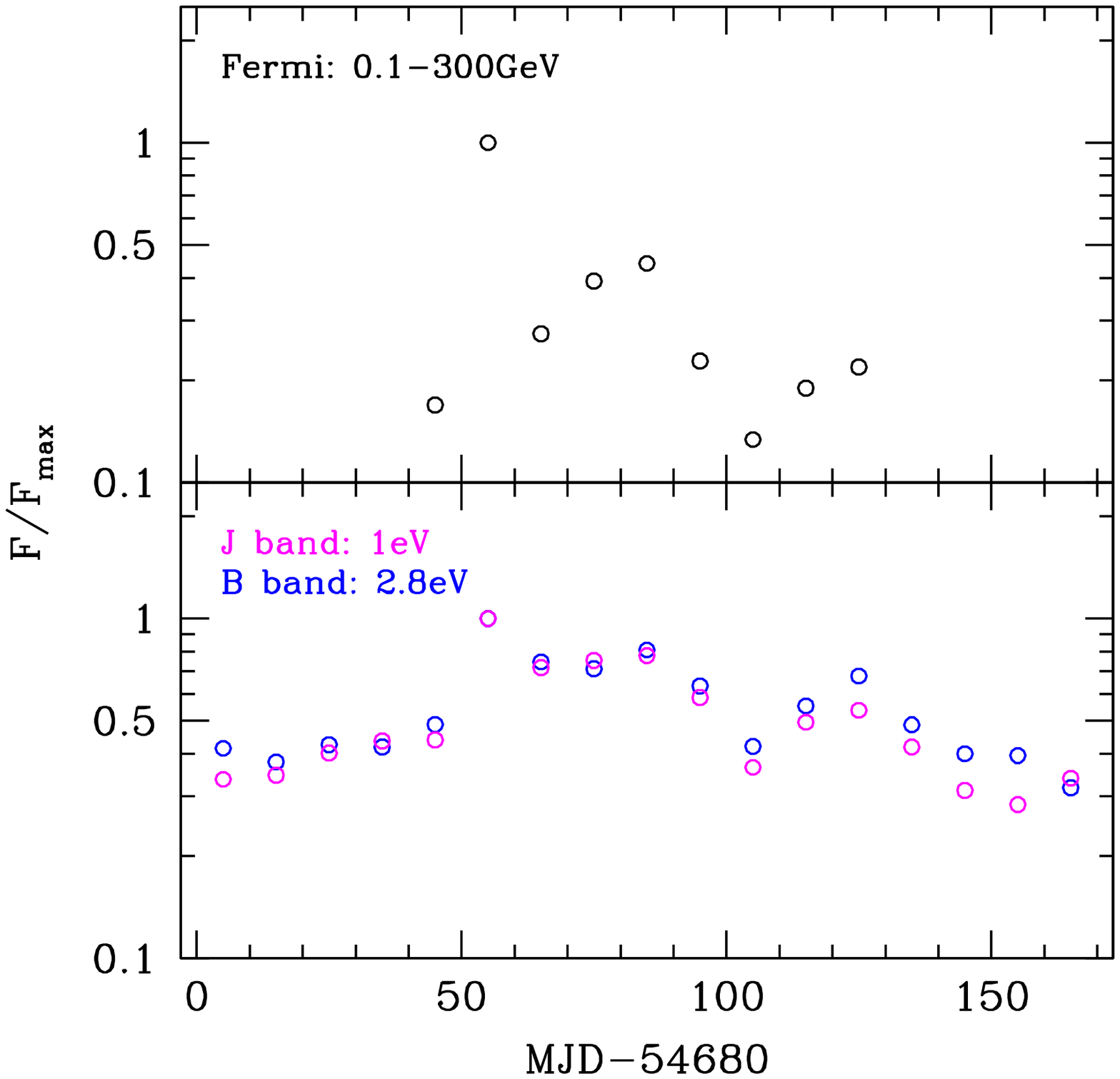}
\hfill
\includegraphics[width=0.68\linewidth]{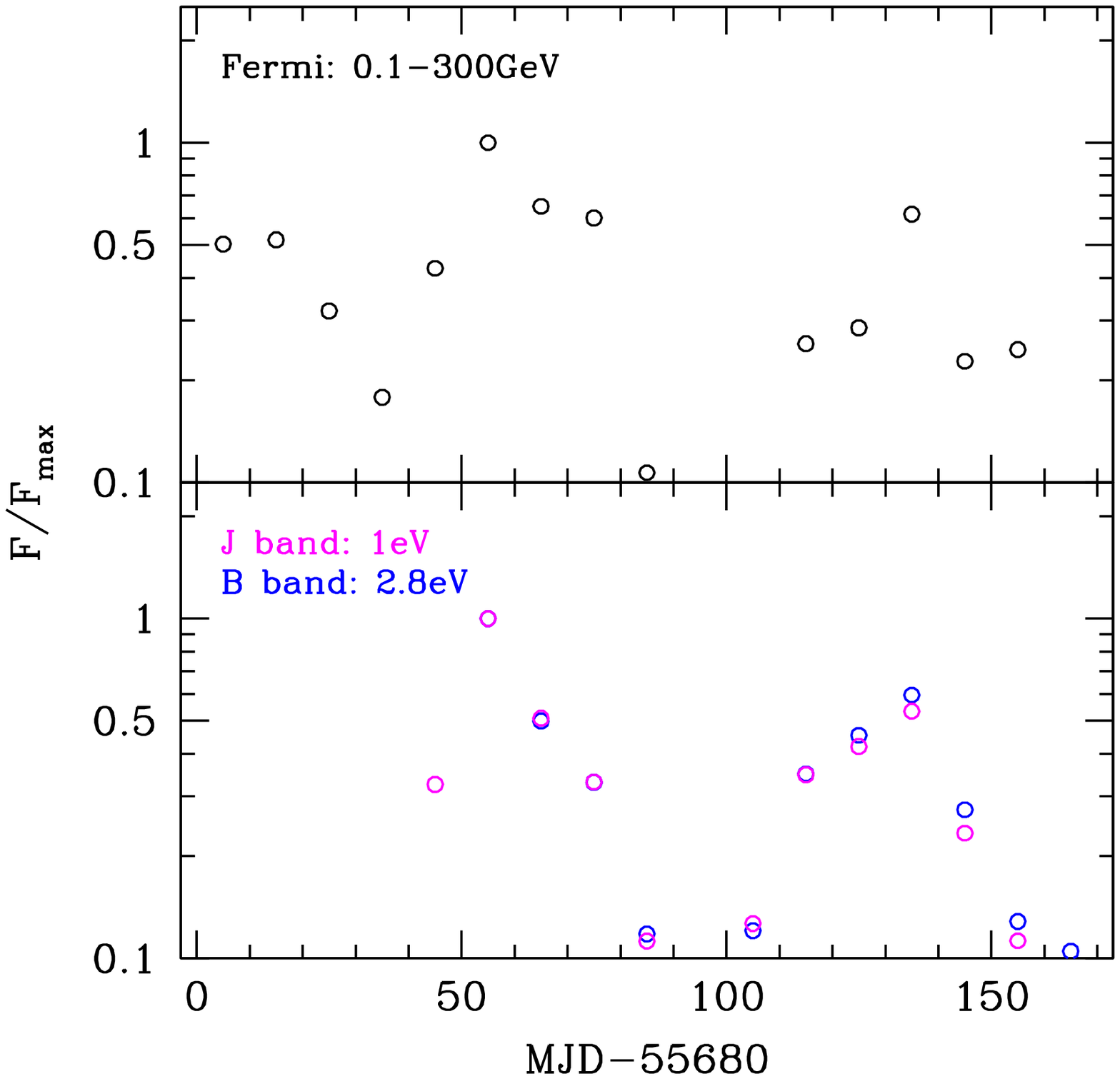}
\hfill

\end{SCfigure*}

Since the optical emission is generally accepted to be
produced by synchrotron emission \cite{urry_mushotzky:1982},
it is sensible to postulate that if the flare is caused by a change of
the magnetic field, it may not have a direct effect on the higher energy emission.
As a turbulent dynamo effect, 
it is known that magnetic field can be amplified by astrophysical shocks beyond simple
shock compression. This has been proved both numerically \cite{guo_2012:mhd_bampli}
and analytically \cite{fraschetti_2013:analytical_bampli}, 
and has been applied to explain
observations of supernovae remnants (SNRs) \cite{parizot_2006:snr_bampli} 
and \gray bursts (GRBs) \cite{mizuno_2011:mhd_bampli}.
If this kind of amplification is also at work in the shocks in the relativistic jets 
which presumably cause the blazar flares, it can be expected to explain the orphan optical flares.

\cite{chatterjee_2013:0208_implication} studied these anomalous flares of PKS 0208-512 
with non-time-dependent modeling. However, since blazar flares are naturally time-dependent
phenomena, it is important to account for the timing information with time-dependent modeling.
We will briefly describe the data handling in $\S$ \ref{observe}.
The comparison between time-dependent modeling and the observation 
are presented in $\S$ \ref{results}, followed by discussion in $\S$ \ref{discussion}.

\section{Multiwavelength data of PKS 0208-512}
\label{observe}

\cite{chatterjee_2013:0208_opticalonly} presented the multiwavelength light curves of PKS 0208-512.
Here we follow similar procedures, but keep the spectral index variable, 
and process the data in 10 day time bins. All the \gray data
shown have test statistic larger than 25, which is comparable to 5$\sigma$ detections.
The optical-infrared data are averages of the daily flux within 10 day bins.
The high- and low- state SEDs from flares 1 and 2 are shown with the simulation results in 
Fig.\,\ref{fig:sscb}-\ref{fig:ecacc}.
In Fig.\,\ref{fig:obs_lc} we show the light curves of flares 1 and 3 as identified by
\cite{chatterjee_2013:0208_opticalonly}. The light curves in optical and \grays show striking similarity
with no apparent delay between the peaks in different energy.
Flare 2 is not detected significantly in \grays in 10 day bins. We show its optical
light curves together with the simulation results.

The \xray data is obtained through the \swift-XRT data products generator
\cite{evans_2009:swiftxrt_web}, and fitted with the \xray spectral fitting package XSPEC.
During the time interval of interet (MJD 55190-55200), \swift-XRT only observed the source
for one 1351s period on MJD 55195.

\section{Modeling Results}
\label{results}

We use the time-dependent inhomogeneous blazar model built by \cite{chen_etal:2011:multizone_code_mrk421,
chen_etal:2012:424.789} to study the multiwavelength data set of PKS 0208-512. 
This model takes a axisymmetric cylindrical geometry. The volume is divided
into many zones in radial and longitudinal directions to account for the inhomogeneity.
This inhomogeneous blob travels relativistically in the AGN frame, and encounters
a stationary shock structure. In the blob frame, it is the shock that travels through
the blob and causes a change in the plasma condition, hence initiating the flare.
For simplicity, this shock is treated as a flat structure.
Monte Carlo method is used for
the radiative transfer, so that all the light travel time effects (LTTEs) are taken into account. 
Fokker-Planck equation is used to follow the evolution of electrons, where synchrotron and 
inverse Compton (IC) cooling, 
as well as stochastic particle accleration and particle escape are present.
The acceleration process is similar to those described by \cite{katarzynski_etal:2006:stochastic}.
It is a result of particle diffusion in momentum space, which mainly represents
the second order Fermi process. The acceleration time scale is assumed to be
independent of particle energy, while its spatial and time variations are treated manually, if any.

\subsection{Pure SSC scenario}
\label{ssc}
\subsubsection{Brief change of magnetic field}
\label{ssc:B}

\begin{figure*}
\centering
\includegraphics[width=0.33\linewidth]{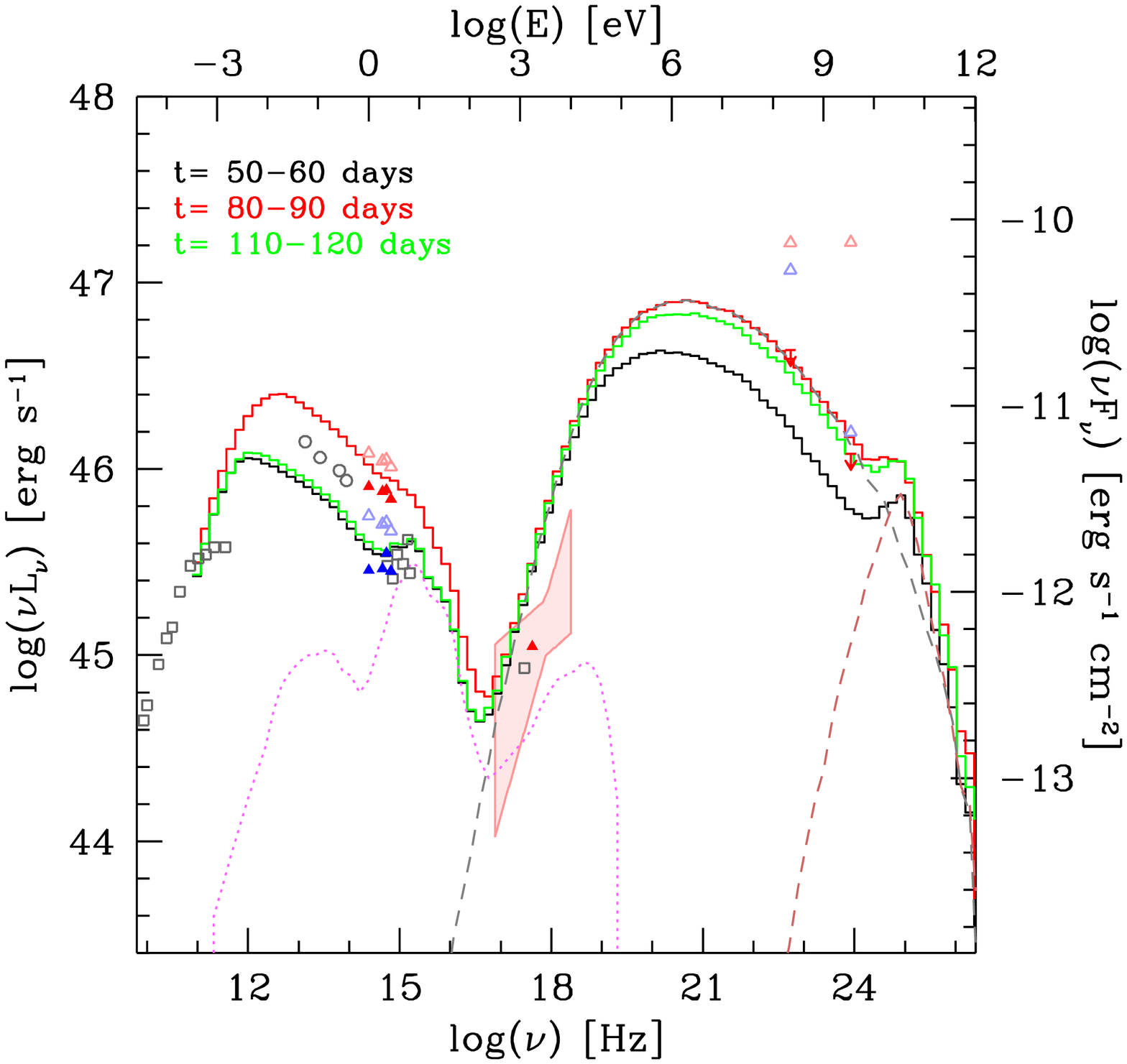}
\hfill
\includegraphics[width=0.33\linewidth]{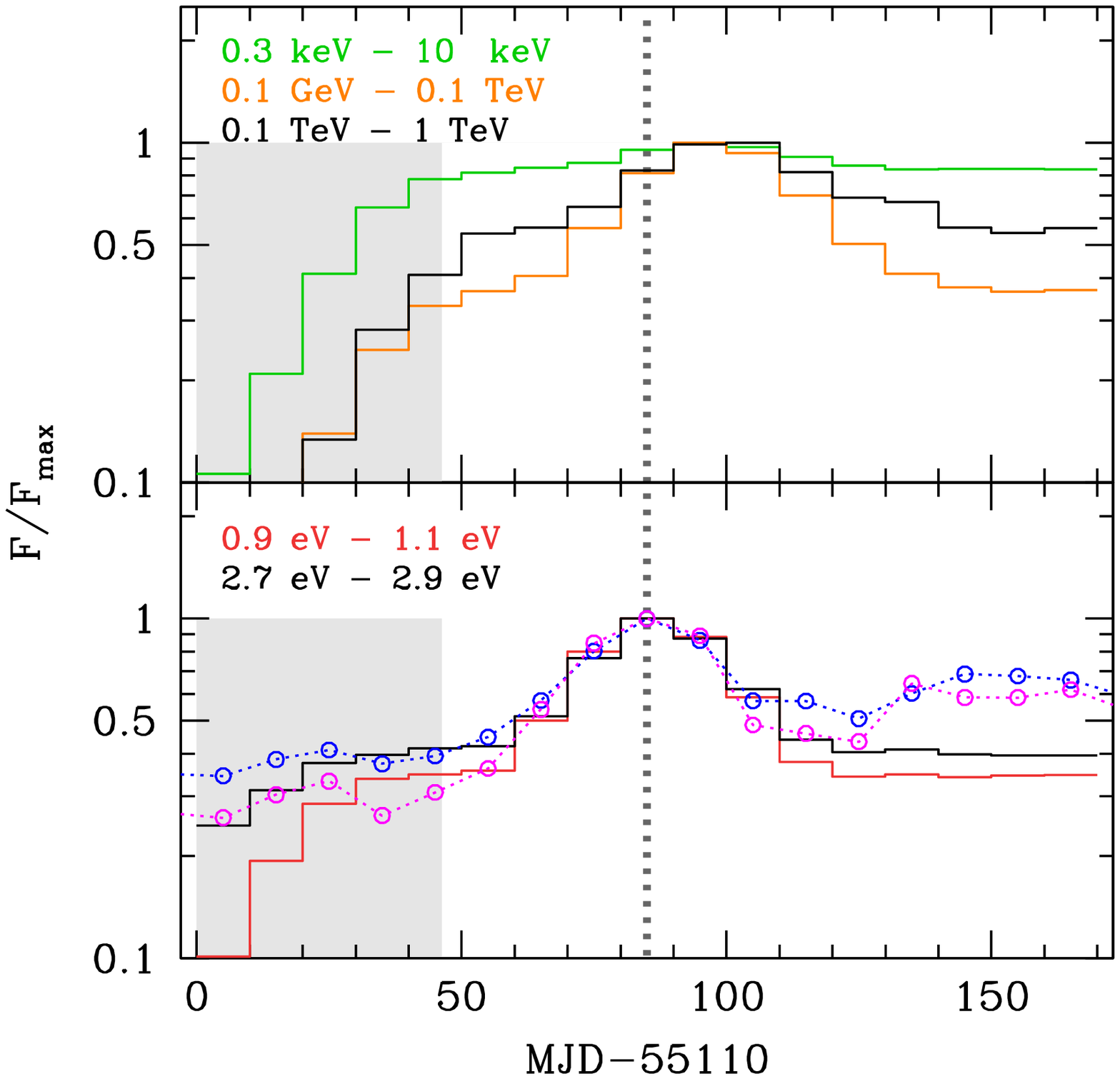}
\hfill
\includegraphics[width=0.33\linewidth]{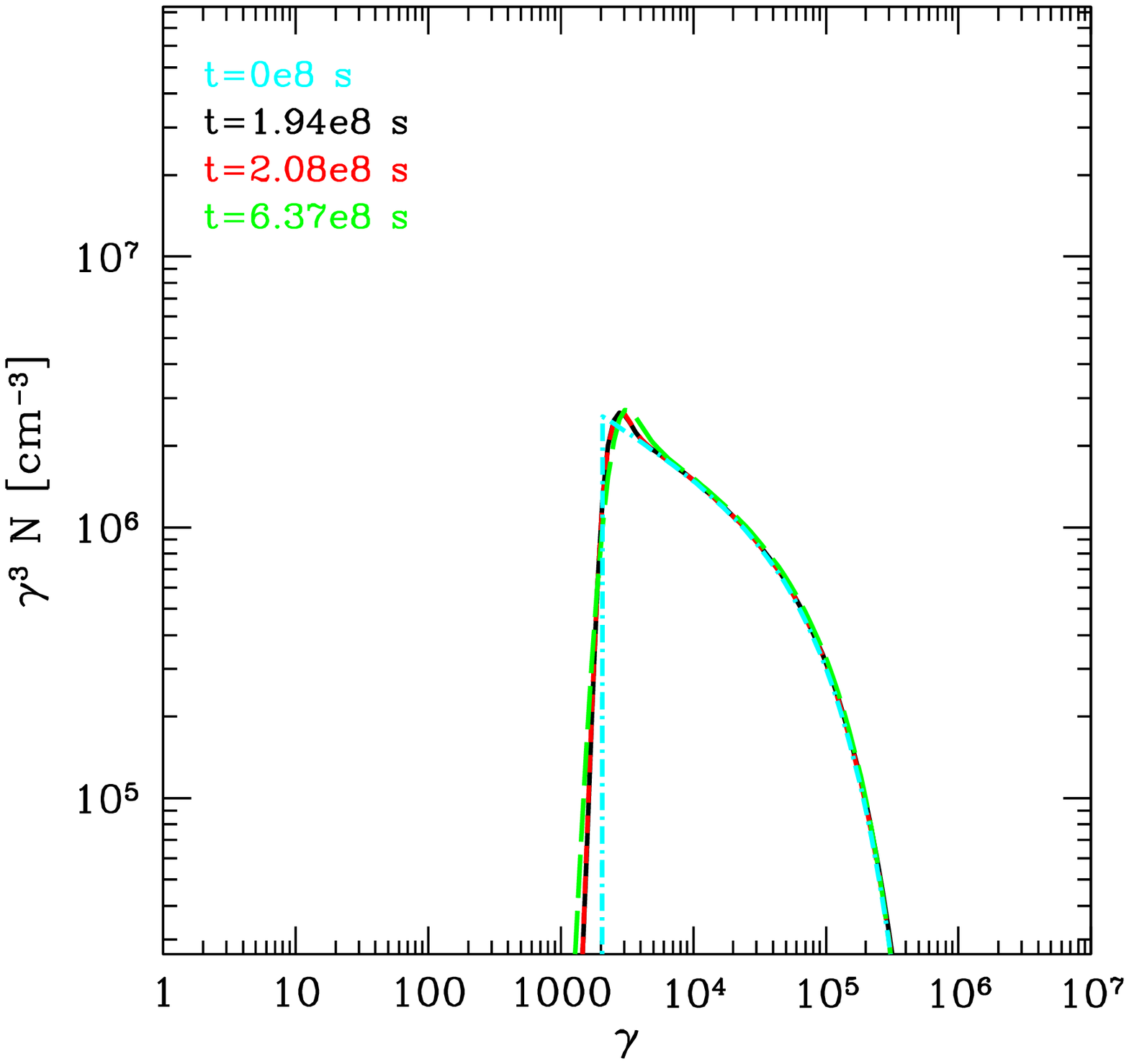}
\caption{The SEDs (left), light curves (middle) and electron distributions (right) 
in the pure Synchrotron Self-Compton (SSC) scenario with
change of the magnetic field as the cause of the flare. 
{\bf Left:} The colored triangles are simultaneous data 
points from SMARTS, \swift and \fermi, 
with red being the high states and blue being the low states. The open triangles are from the 
flare 1 identified by \cite{chatterjee_2013:0208_opticalonly}, 
while the filled triangles are from flare 2. 
The open blue triangles are from MJD 54720-54730;
the open red triangles are from MJD 54730-54740; 
the filled blue triangles are from MJD 55160-55170;
the filled red triangles and bow-tie (SMARTS and \swift), as well as the red 
\fermi upper-limits are from MJD 55190-55200.
The grey squares include \swift, \planck and ground based radio data 
in November of 2009 reported by \cite{giommi_2012:planck}.
The gray circles are \wise data taken in MJD 55367-55369 (around June 23, 2010).
The three histograms show the SEDs before, during, and after the peak of a flare, 
with legends in corresponding colors showing the simulation time in the observer's frame.
The dotted magenta line is the median thermal emission from radio loud quasars \cite{elvis_etal94_qsr_sed} 
scaled according to the observed UVOT flux of this source. It is added to the simulated SEDs 
in the post-processing as a steady component. The dashed red line is the isolated second order SSC emission 
during the peak of the flare,
while the dashed grey line shows the first order SSC.
{\bf Middle:} In the bottom panel the open circles show the 10-day-averaged optical light curves 
in B (blue) and J (magenta) bands starting from MJD 55110. The data points are connected by dotted lines to guide the eyes. The histograms show two simulated synchrotron light curves
at similar frequencies. In the upper panel three simulated IC light curves are shown, with green, orange,
and black solid lines representing the energy bands in \xray, \fermi \gray, and very high energy (VHE) \gray. The shaded gray areas
mark the phase when the simulation is still in setup phase. The vertical dotted line marks the peak of
the synchrotron light curves.
{\bf Right:} The electron distributions in the front-center zone. The simulation time shown are 
based in the frame of the emitting blob.
}
\label{fig:sscb}       
\end{figure*}

We begin the investigation with a pure SSC scenario. In this case, five key parameters (magnetic field $B$, 
electron density $n_{e}$, volume length $Z$ (or radius $R$), beaming Doppler factor $\delta$,
Lorentz factor of the injected low energy particle $\gamma_{inj}$) are constrained
by 5 observables (synchrotron and IC peak frequencies $\nu_{sy}, \nu_{ic}$, synchrotron and IC 
apparent luminosity $L_{sy}, L_{ic}$, and variability time scale $t_{var}$). 
The parameters used are summarized in Table \ref{tab-1}.

The flare is assumed to be caused by an increase of magnetic field energy density
(by a factor of 20) immediately downstream of
the stationary shock. The thickness of the region with increased magnetic field is $1/10$
of the length of the emitting blob. The same thickness is used in subsequent cases.

The results (SEDs, light curves, and electron distributions) of the modeling are shown in 
Fig.\,\ref{fig:sscb}. In both optical and \gray bands, the blazar shows flaring behaviors, although the \gray
flares seem to be more smoothed and spread out. The peak of the \fermi \gray light curve has a prominent
delay compared to the peak of the optical light curve. The delay is roughly 10 days in this case, 
which is about 1/5 of the variability time scale.
Both the delay and smoothing of the \gray light curve are caused by the internal LTTE of the SSC emission. 
The inconsistency between this delay and the lack of time delays in observations such as those shown in 
Fig.\,\ref{fig:obs_lc} indicate that the SSC with magnetic field strength change scenario does not explain
the correlated flares in FSRQs for the geometries studied here. Neither does it explain the optical only flares obviously.

If the flare is not caused by a change in the magnetic field strength, but by some intrinsic change to the electrons,
the time delay property might be different. However, in this constrained SSC scenario, the deduced strength 
of magnetic field is only $0.4mG$ in the quiescent state, 
making the cooling of the electrons quite slow compared to the
flare time scale (see Fig.\,\ref{fig:sscb} right). 
One result of this slow cooling is that any change of the electrons (e.g. injection of
new high energy electrons) will not be able to cool down to the quiescent state soon. The resulted light curves
would have sharp rising time, but much longer decay time.

Another interesting implication of Fig.\,\ref{fig:sscb} is the presence of second order SSC emission above
tens of GeV. Despite being largely suppressed by Klein-Nishina (KN) effect, the second order SSC does show up
as an additional bump in the SED.
Since the SEDs of some high redshift blazars such as 
3C 279 \cite{albert_etal:2008:3C279_magic_detection} and PKS 1424+240 \cite{furniss_2013:1424_redshift} 
appear to have an up-curving shape at VHE after
the correction of extragalactic background light (EBL) absorption, 
it is tempting to explain those SEDs with the second order SSC emission.
However, PKS 1424+240 is classified as an intermediate-frequency peaked BL Lac (IBL) or 
high-frequency peaked BL Lac (HBL). The higher energy synchrotron peak of this source means 
it is very unlikely
to show significant second order SSC emission due to strong KN suppression.
Since these different types of blazars show similar up-curving SED only after EBL de-absorption,
it is more natural to explain these curves as a result of the uncertainty in the EBL models used.

\begin{figure*}
\centering
\includegraphics[width=0.33\linewidth]{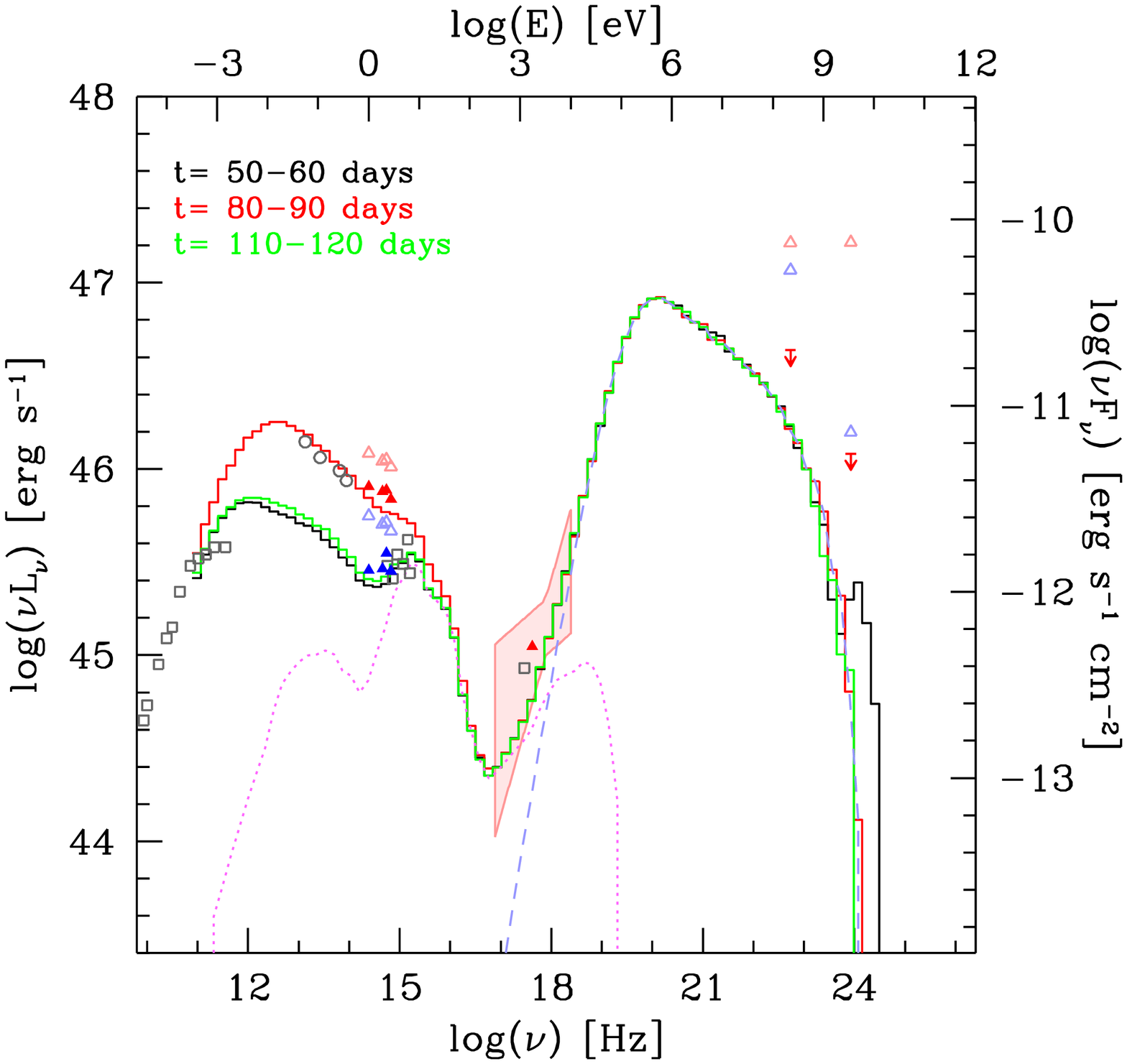}
\hfill
\includegraphics[width=0.33\linewidth]{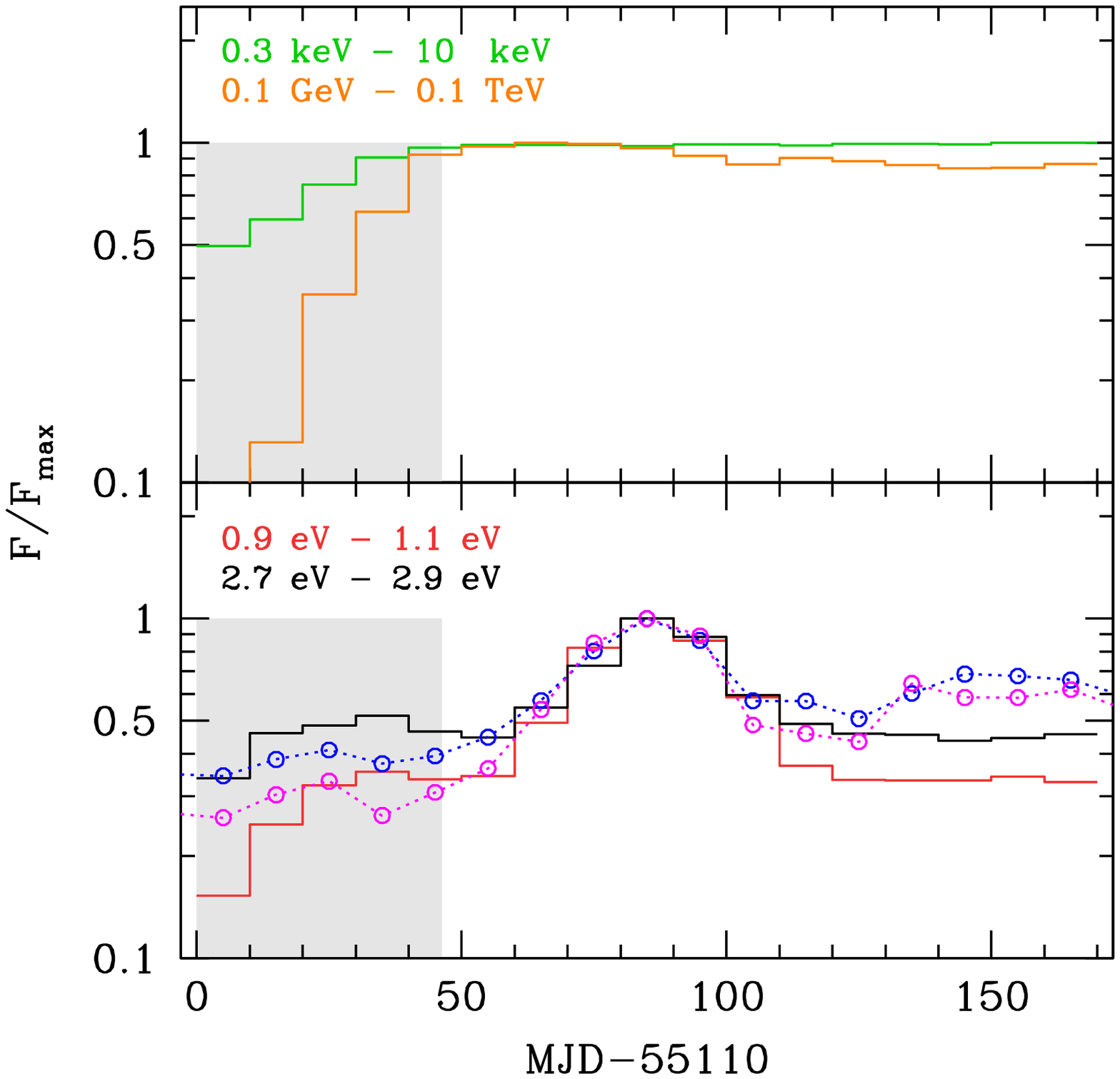}
\hfill
\includegraphics[width=0.33\linewidth]{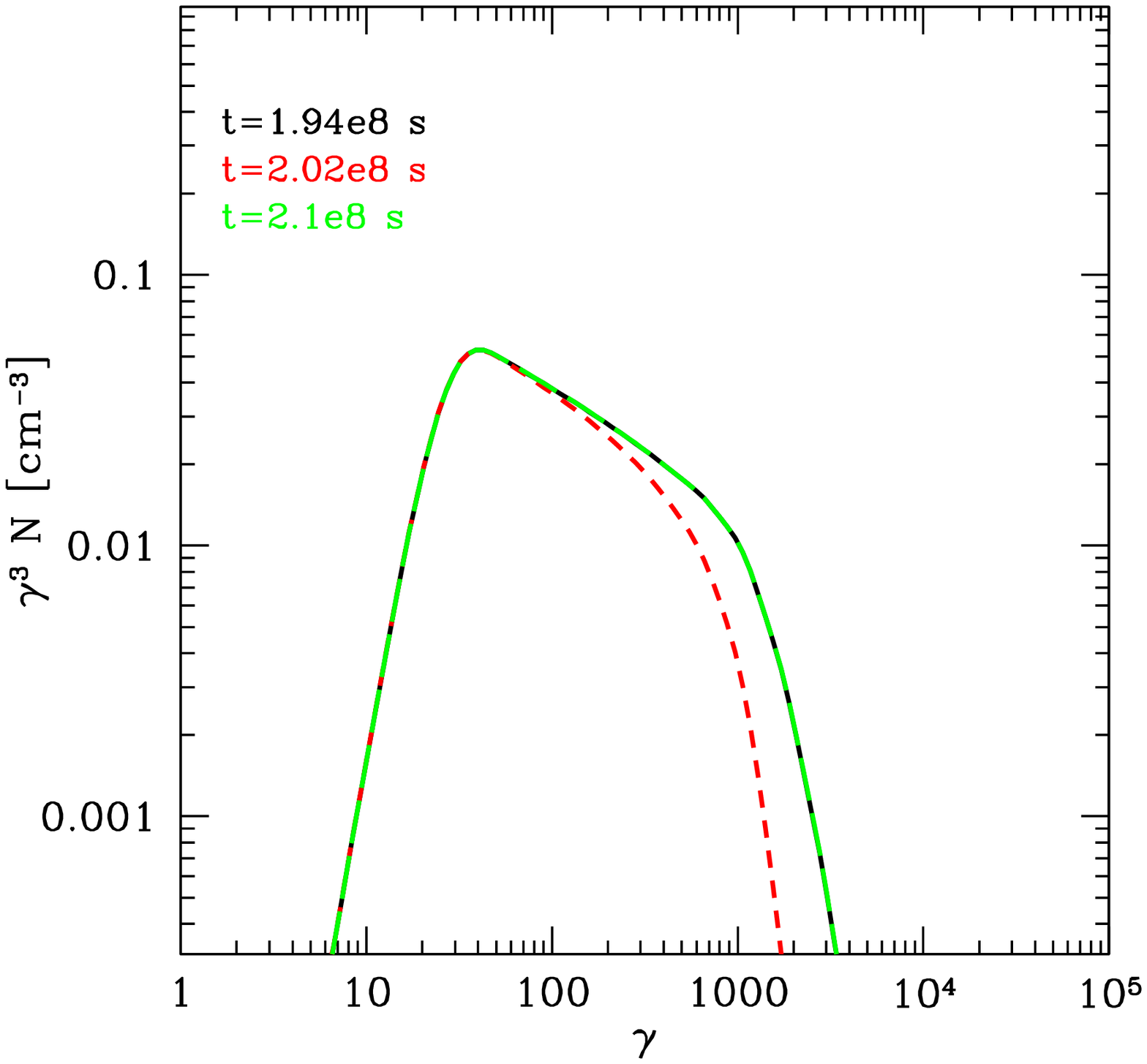}
\caption{The SEDs (left), light curves (middle) and electron distributions (right) of the EC/dusty torus 
scenario with change of the magnetic field as the cause of the flare. 
The data points are similar to those in \ref{fig:sscb}, except that in the SED the blue dashed line
shows the EC component, and in the light curves there is no VHE \gray because of zero flux.}
\label{fig:ecb}       
\end{figure*}

\subsection{Dusty torus EC scenarios}
\label{ec}

With the difficulty met in a pure SSC scenario, we turn to external Compton (EC) 
process as the mechanism responsible for
the \gray emission. We chose the emission from the dusty torus as the source of the external emission.
The detection of FSRQs by imaging atmospheric Cherenkov telescopes (IACT) above 100GeV 
\cite{albert_etal:2008:3C279_magic_detection,alksic_2011:1222_magic_detection:730.8,hess_2013:1510} 
locates the \gray emission site to be outside of the broad line region (BLR), otherwise the \grays would
not be able to escape the $\gamma-\gamma$ absorption by the BLR photons \cite{tavecchio_2012:flat_blr}.
Compared to the pure SSC scenario, the EC scenario has one additional parameter, 
that is the energy density of the 
external photons. This parameter is connected to luminosity of the quasar thermal emission. However, the
large uncertainty in the torus radius and covering factor means it is poorly constrained. So in the EC
scenario, we fix the bulk Lorentz factor $\Gamma$ 
(and hence the Doppler factor $\delta$, because we always use the line of sight to jet axis
angle $\theta=1/\Gamma$) to 40, which is the value used in the SSC scenario. 
This is also close to the largest value determined in VLBI observation of quasar
jets \cite{jorstad_etal:2005:VLBA}.

\subsubsection{Brief change of magnetic field}
\label{ec:B}

Fig.\,\ref{fig:ecb} shows the results of the EC case with the brief change of magnetic field strength as
the cause of the flare. The magnetic field energy density is increased by a factor of 27 downstream of the shock. 
The results show that the optical emission is strongly variable, but the IC 
emissions remain relatively quiet. This is a fair reproduction of the orphan optical flare observed.
However, this alone can not yet explain the co-existence of the orphan optical flares and the optical/\gray 
correlated flares.
There must be other processes that are involved in the correlated flares.

\begin{table}
\centering
\caption{The parameters used in the quiescent state of the simulations. 
Observation angle is chosen to be $1/\Gamma$ so that
the Doppler factor $\delta$ always equals the bulk Lorentz factor $\Gamma$.
The volume radius $R=3Z/4$ in all cases.
Particle acceleration time scale and particle escape time scale have the ratio $t_{acc}/t_{esc}=6.5$
except during the flare of the case described in $\S$\ref{ec:acc}}
\label{tab-1}       
\begin{tabular}{lll}
\hline
  & SSC & EC \\\hline
B(G) & $4\times10^{-4}$ & 2 \\
$\delta$ & 40 & 40 \\
$\gamma_{inj}$ & $2.4\times10^{3}$ & 20 \\
$Z(cm)$ & $2.4\times10^{18}$ & $2.4\times10^{18}$ \\
$n_e(cm^{-3})$ & 0.3 & $8\times10^{-5}$ \\
$t_{acc}(Z/c)$ & 120 & $1.3\times10^{-4}$ \\\hline
\end{tabular}
\end{table}
\begin{figure*}
\centering
\includegraphics[width=0.33\linewidth]{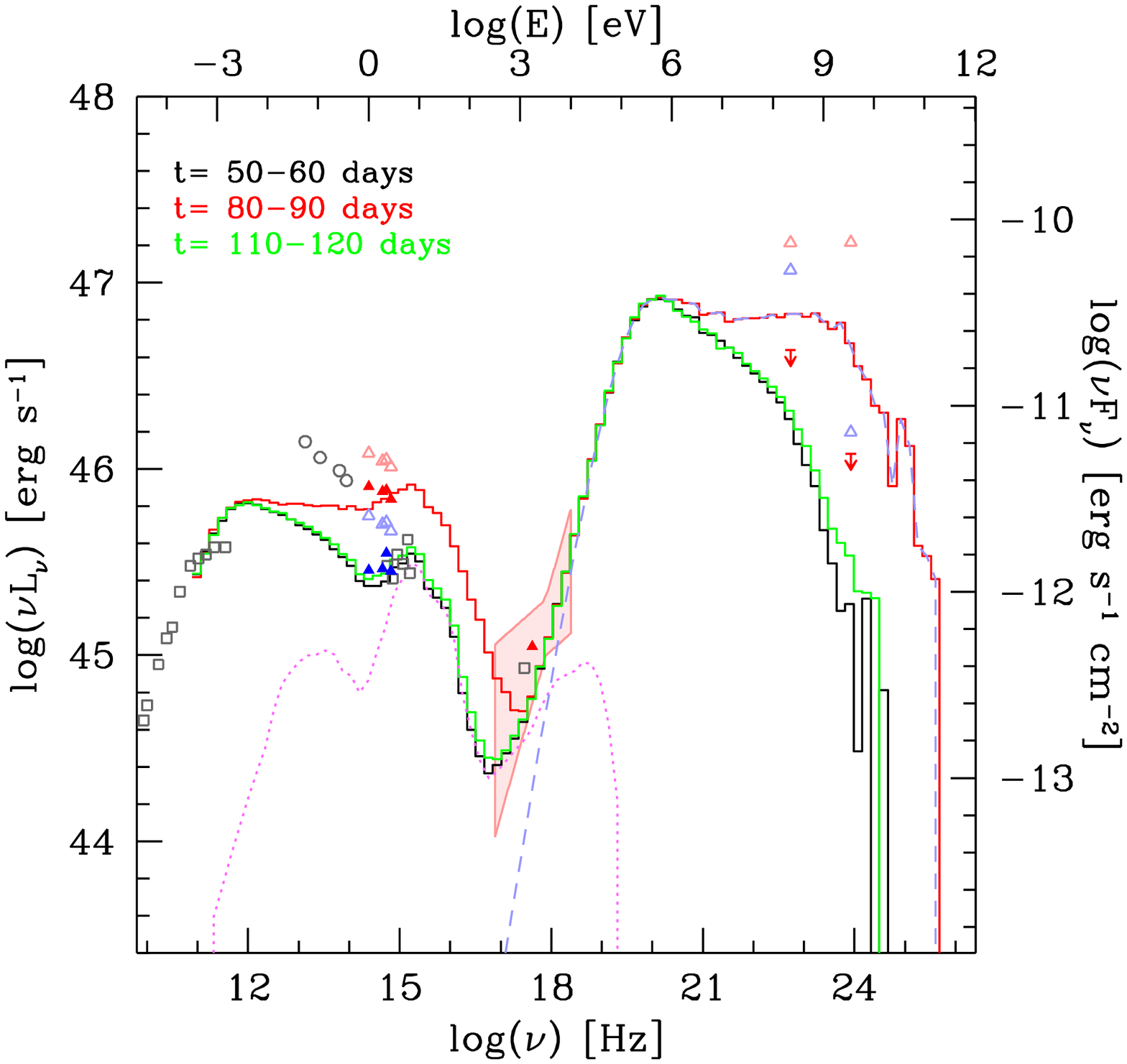}
\hfill
\includegraphics[width=0.33\linewidth]{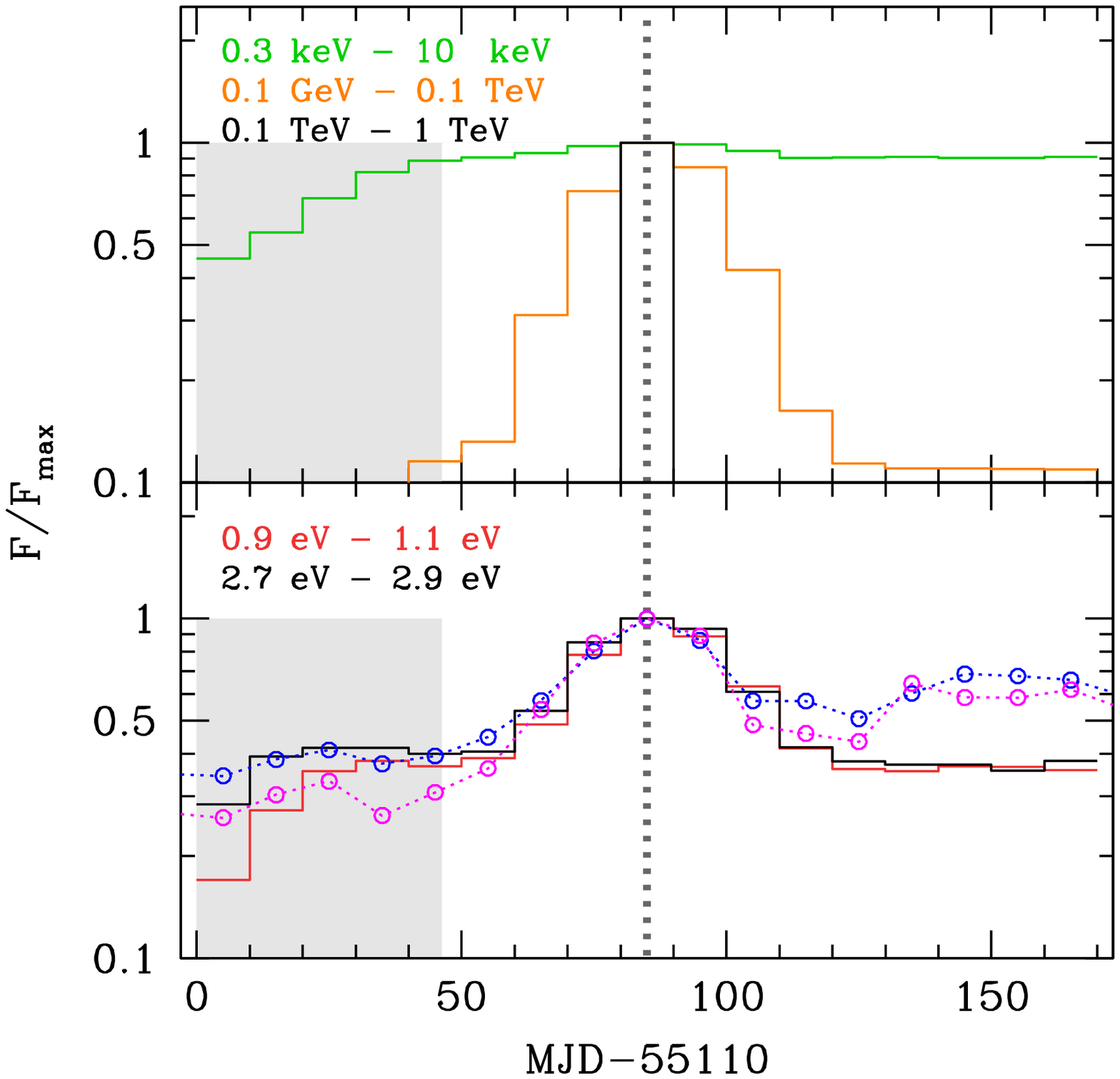}
\hfill
\includegraphics[width=0.33\linewidth]{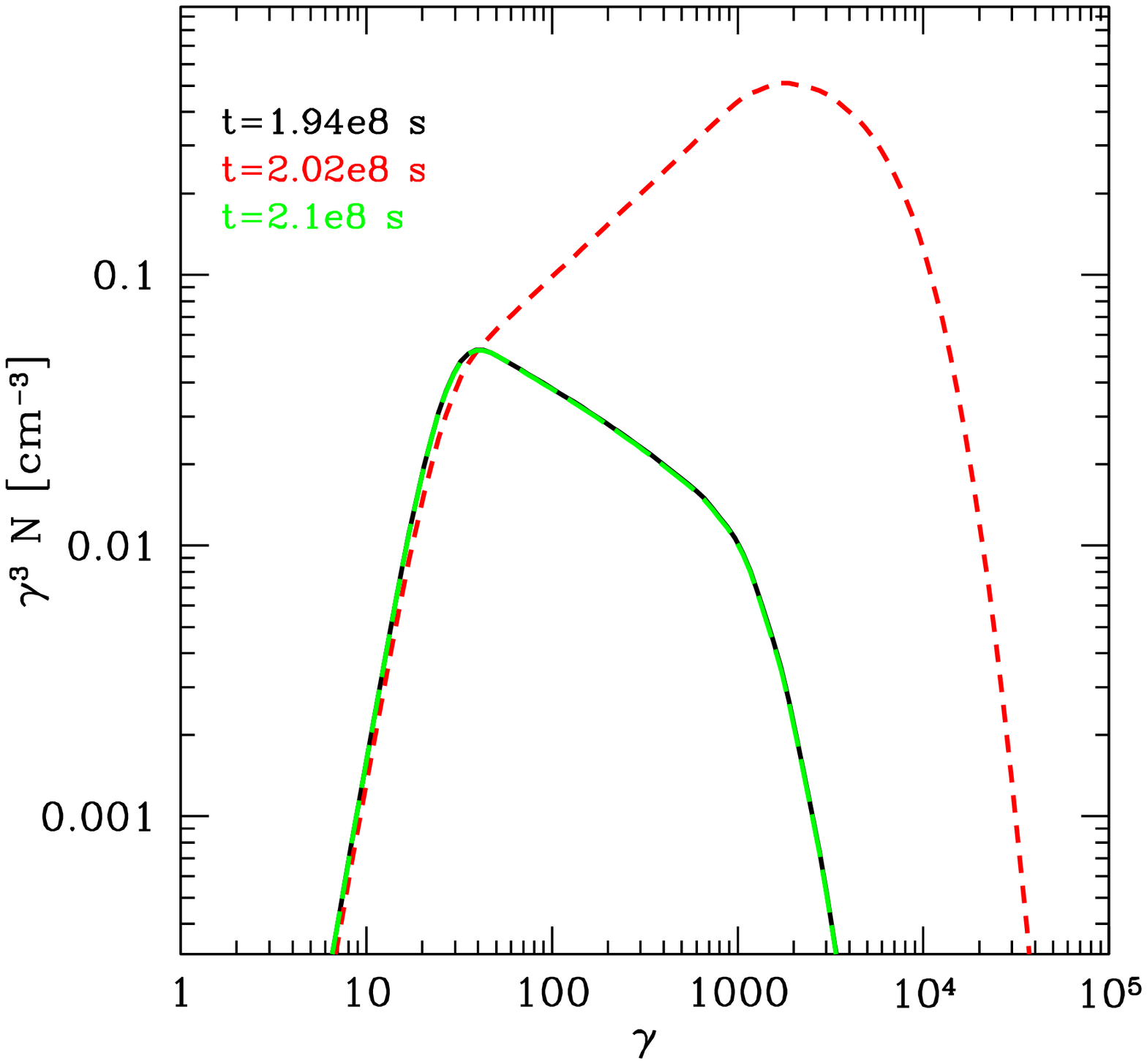}
\caption{The SEDs (left), light curves (middle) and electron distributions (right) of the EC/dusty torus 
scenario with change of particle acceleration efficiency as the cause of the flare. 
The data points are similar to those in \ref{fig:sscb},  except that in the SED the blue dashed line
shows the EC component.}
\label{fig:ecacc}       
\end{figure*}
\subsubsection{Brief change of acceleration efficiency}
\label{ec:acc}

All current theories of magnetic field amplification involve the generation of strong turbulence.
These strong turbulence can be expected to cause strong stochastic particle acceleration.
Keeping this in mind, we also evaluate the effect of changing the particle acceleration efficiency
($t_{acc}$ shortened by a factor of 2.2)
in a thin shocked layer. The results are shown in Fig.\,\ref{fig:ecacc}.

As seen in Fig.\,\ref{fig:ecacc}, the optical and \gray activities are well correlated, with no apparent delay. This proves that in the same
EC scenario, optical/\gray correlated flares are also possible. 

Another prominent feature of the results
is the spectral hardening during the flare, both in synchrotron and IC emission. These
spectral hardening is observed in \fermi monitored FSRQs
\cite{abdo_etal:2010:spectral_properties_of_blazars}. But in the narrow optical and infrared band, 
observation reveals a redder when brighter trend \cite{bonning_2012:smarts_blazar},
most likely due to the contamination from the quasar thermal emission. 
This is also consistent with the simulated
light curves in Fig.\,\ref{fig:ecacc}. Observations in a wider far-infrared band can better
resolve the spectral behavior of the synchrotron component
(see \cite{nalewajko_2012:1510_herschel}, where \herschel sees a harder when brighter trend 
in PKS 1510-089).

Moreover, the simulation indicates that the SED extends to higher energy during the flare, causing a VHE
\gray flare that is observable to ground based Cherenkov telescopes. This is because when the particle
acceleration efficiency increases, the electron distribution does not only become harder, but also reaches
larger maximum electron energy (see Fig.\,\ref{fig:ecacc} right). For the same reason, the synchrotron
emission also extends to higher energy, causing an ultrviolet flare and the intrusion of the synchrotron emission to 
the soft \xray energy. The simultaneous multi-wavelength data we have for PKS 0208-512 are not sufficient 
to confirm these ultrviolet, \xray and VHE behaviors. However, 3 other FSRQs have already been observed by 
Cherenkov Telescopes during flaring states 
\cite{albert_etal:2008:3C279_magic_detection,alksic_2011:1222_magic_detection:730.8,hess_2013:1510}. 
The upcurving \xray SED was recently observed in one FSRQ (PNM J2345-1555) 
during a major flare,
accompanied by spectral hardening in GeV \gray \cite{gg_2013:2345_redblue}.
Our simulation suggests that these \xray softening, ultraviolet/VHE \gray flaring, 
and far-infrared/\gray spectral hardening
are related features of correlated optical/\gray flares. We suggest that strong spectral softening
in \xray monitoring and ultraviolet flares can be used as triggers for IACT observations in search of VHE emission from FSRQs.

\section{Discussion}
\label{discussion}
With our time dependent inhomogeneous blazar model, 
we studied blazar flares caused by changes of magnetic field and particle acceleration efficiency, 
in both SSC and EC scenarios. SSC was disfavored mainly because the model predicts a delay of the \gray
emission, while observations show on average such delays are not prefered in FSRQs. In the EC scenario,
change of particle acceleration efficiency can explain, during \gray flares of FSRQs, 
the spectral hardening of \grays, the rare detections at VHE \grays, 
and the upcurving \xray SEDs. Based on this result, we recommend to include
the softness of \xray spectrum as an indicator in the search of VHE emission from FSRQs by IACTs.

We reproduced both orphan optical and optical/\gray correlated flares with the same quiescent-state emission
setup in the EC scenario. The difference lies in the causes of the flares, i.e. whether the flare is caused
by a change of magnetic field strength or particle acceleration efficiency. This difference can be
attributed to the allocation of shock energy between magnetization and turbulence. This allocation further
depends on the initial orientation of the magnetic field. \cite{mizuno_2011:mhd_bampli} showed with MHD 
simulation that with magnetic field perpendicular to the shock flow, more energy is partitioned to the
magnetic field, because the initial magnetic field is already compressed by the shock before any further
turbulence induced amplification. 
We argue that this corresponds to the orphan optical flares, and the parallel magnetic field case
corresponds to the optical/\gray correlated flares.
At the same time, this perpendicular magnetic field case has a larger portion of
compressed ordered magnetic field downstream of the shock, 
so we can expect the emission to have stronger optical polarization. Based on this interpretation,
the increase in polarization will be 
accompanied by the increase in optical flux while its correlation with \gray flux will be weak.
Hence the degree of polarization should correlate with the optical/\gray flux ratio. Whether this correlation
exists awaits further confirmation from observations.



\subsection{External Radiation Field}

The source of external radiation in this study is assumed to be the dusty torus. However, the results in
this study is qualitatively applicable to most other kinds of external radiation field. 
The size of the emission blob deduced from this study is comparable to the size of the dusty torus 
($R_{ir}=3.5\times10^{18}cm$). 
This implies that while the blob travels on the time scale under consideration,
it may experience considerable change of the external radiation energy density.
This change is not included in our current simulations.
We can expect that the changes in the external field will lead to \gray variations which do not have 
optical counterparts. Since the occurrence of such variation in FSRQs is still unclear, we leave these
considerations to future work.

\subsection{Particle Escape}

The rate of particle escape required for PKS 0208-512 in our EC scenario is very fast ($t_{esc}=2\times10^{-5}R/c$).
This is required because of the similarly fast particle acceleration that is needed to balance the 
efficient radiative cooling in this case. This particle escape is too fast to be explained by 
particle streaming out of the emission region. 

One possible explanation is that the particles are
accelerated in much smaller turbulent cells 
(see \cite{marscher_2013:fermi_symp_turb} for discussion of the idea of turbulent cells) that are
distributed across the emission region. So the `escape' used in our model does not represent
escape from the emission region, but rather escape from the accelerators. 
For example, if the size of the turbulent cells is $R_{cell}=10^{-5}R/c$,
$t_{esc}=2R_{cell}/c$.
These escaped particles
are no longer being accelerated, but they still contribute to both synchrotron and IC emission.
Let us assume the external radiation and its associated IC cooling are homogenous.
As particles are being cooled in the larger volume and being injected from the cells, the particle number outside of the cells $N(\gamma)$ reaches a steady state
$(p_{sec}-2)N(\gamma)/t_{cool}=N_{cell}(\gamma)/t_{esc}$, at sections where it can be described by a power-law with slope $p_{sec}$.
With particle number inside the cells $N_{cell}(\gamma)$ having a power-law distribution,
$N(\gamma)$ will form a broken power-law distribution
with the break energy equal to $\gamma_{inj}$. Below the break $p_{sec}=2$,
while above the break $p_{sec}=p$, where $p-1$ is the power-law slope of $N_{cell}(\gamma)$.
There is an energy $\gamma_{eq}$ where $N_{cell}(\gamma)=N(\gamma)$. For $p\approx3$, this happens close to the energy where $t_{cool}=t_{esc}$ ($\sim t_{acc}/5$).
Below $\gamma_{eq}$ the number of escaped particles
is greater than that of the unescaped particles
(at energy $\gamma>\gamma_{inj}$, $N(\gamma)=\frac{\gamma_{eq}}{\gamma}N_{cell}(\gamma)$).
If the magnetic field is also homogeneous, 
both the synchrotron and IC emission of the escaped particles will dominate. 
Under this scenario, the particle density $n_e$ listed in Table \ref{tab-1}
($n_e=8\times10^{-5} cm^{-3}$) represents 
the escaped particle density above
$\gamma_{inj}$, while the total escaped non-thermal particle density 
$n_{total}\sim n_e(p-1)\gamma_{inj}^2=0.07 cm^{-3}$ when $p=3.2$.
In the whole volume the unescaped particles 
have averaged density of about
$\frac{p-1}{p-2}\frac{\gamma_{inj}}{\gamma_{eq}}n_e=6\times10^{-7} cm^{-3}$
when $\gamma_{eq}\sim5\gamma_{max}\approx 5000$.
The time dependent behavior of the escaped particles above $\gamma_{inj}$ will 
follow those of the unescaped ones. With a change of the normalization, and 
 the power-law slope by 1, 
the emission of the unescaped particles can represent those of the escaped ones.

For the sake of universality, in this work we do not directly model the escaped particle population,
because there are other possible explanations.
For example the energy independence of the acceleration time scale used in this study may be a poor approximation,
and a more realistic treatment of the energy dependence of the acceleration may result in very different
particle escape rate to maintain the same electron distribution.

\section*{Acknowledgements}

XC and MP acknowledge support by the Helmholtz Alliance for Astroparticle Physics HAP funded by the Initiative and Networking Fund of the Helmholtz Association.


\bibliography{refs_all}
%
%
%
%
\end{document}